\documentclass[twocolumn,showpacs,preprintnumbers,amsmath,amssymb,floatfix]{revtex4}

\usepackage{graphicx}
\usepackage{dcolumn}
\usepackage{bm}
\usepackage{psfrag}

\begin{document}

\preprint{APS/123-QED}

\title{Fluctuation induced effects in confined electrolyte solutions}

\author{T. Brandes}%
\email{brandes@dirac.phy.umist.ac.uk}
\author{L. Lue}
\altaffiliation[]{Department of Chemical Engineering}
\email{leo.lue@umist.ac.uk}
\affiliation{%
Department of Physics and
Department of Chemical Engineering \\
UMIST \\
PO Box 88 \\
Manchester M60 1QD \\
United Kingdom 
}%


\date{\today}

\begin{abstract}
We examine electrolyte systems confined between two parallel, grounded
metal plates using a field-theoretic approach truncated at one-loop
order.
For symmetric electrolytes, the system is locally neutral, and the
electric potential between the plates is everywhere zero, in agreement
with the predictions of the Poisson-Boltzmann equation.  The density
distribution of the ions, however, is non-uniform, with a maximum
value at the center of the system and that lies below the bulk
electrolyte density.  This differs from the Poisson-Boltzmann result,
which is uniform and equal to the bulk electrolyte density.
For asymmetric electrolytes, the system is no longer locally neutral,
and there is a non-zero static electric potential.  In addition, the
surface of the metal plates becomes charged.  When the plate
separation is smaller than the screening length, the charge
distribution has a single peak at the center of the system.  As the
separation becomes greater than the screening length, the charge
distribution possesses two peaks with a local minimum at the center of
the system.  These peaks remain localized near the surface of the
plates, while the bulk of the system has nearly zero charge.
When the plates are immersed in a bulk electrolyte solution, an
attractive fluctation induced force develops between them that is
analogous to the Casimir force.  Unlike the Casimir force, however,
this attractive interaction has a finite limit in the zero separation
limit and decays inversely with the separation at large separations.
\end{abstract}

\pacs{Valid PACS appear here}
\maketitle

\section{\label{sec:level1}Introduction}

The electrical double layer near a solid/electrolyte interface occurs
in a wide variety of contexts and plays a crucial role in many
colloidal, electrochemical, and biological systems.  Much of our
understanding of these systems are based on the Poisson-Boltzmann
equation, which has been quite successful
\cite{Gouy_1910,Chapman_1913,Israelachvili_1991}.
This equation is based on a mean-field type approximation,
and, consequently, neglects fluctuation effects which result in
ion-ion correlations.  In many cases, these fluctuations are not
crucial, and only add a qualitative change in the predictions.

In certain situations, however, the contribution of fluctuations are
important and lead to qualitatively different behavior than that
predicted by the Poisson-Boltzmann equation \cite{Levin_2002}.
Examples include the presence of an attractive interaction between two
like-charged colloids \cite{Wu_etal_1999,Moreira_Netz_2001}, the
collapse of polyelectrolytes \cite{Golestanian_etal_1999},
and charge segregation in confined electrolytes
\cite{Yu_etal_1997,Sheng_Tsao_2002}.



Due to the importance of fluctuations, many theoretical methods have
been developed to incorporate fluctuation corrections in electrolyte
systems.  Some of these include liquid-state theory approaches based
on the Ornstein-Zernike equation coupled with an approximation
closure, such as the hypernetted-chain or mean-spherical approximation
\cite{Henderson_Blum_1978,Kjellander_Marcelja_1985,Attard_etal_1988a,Attard_etal_1988b},
density functional theories
\cite{Biben_etal_1998,Boda_etal_1999,Boda_etal_2002}, the modified
Poisson-Boltzmann equation \cite{Outhwaite_etal_1980}, computer
simulation methods
\cite{Torrie_Valleau_1980,Torrie_etal_1982,Boda_etal_1999,Boda_etal_2002},
and field theoretic approaches
\cite{Kholodenko_Beyerlein_1986,Ortner_1999,Netz_Orland_2000b,Brown_Yaffe_2001}.






We examine a pair of grounded, metal plates which are separated by a
distance $L$ and immersed in an electrolyte solution which is
dissolved in a continuum with uniform dielectric constant $\epsilon$.
In this situation, the Poisson-Boltzmann equation predicts that the
system behaves trivially: the ion concentration between the plates is
uniform, and the electric potential is everywhere zero.
For this system, the fluctuation corrections are non-trivial and lead
to some interesting phenomena.  In this case, the ion concentration is
non-uniform, and in some situations can lead to the presence of a
non-zero electric potential and a surface charge on the metal plates.
In addition, we find a fluctuation induced attractive force, analogous
to the Casimir force, between metal plates.

The remainder of the paper is organized as follows.  In Section
\ref{functional:sec}, we quickly review the functional integral
formulation of electrolyte systems.  Then in Section
\ref{examples:sec}, we quickly summarize the main results in
mean-field theory (i.e., the Poisson-Boltzmann equation) for
electrolytes confined between two metal plates and then present field
theory calculations to order one-loop.  Finally, in Section
\ref{conclusions:sec} we summarize the major finding in this work.

\section{Functional integral formulation}

\label{functional:sec}

Consider an electrolyte system that is immersed in a continuous medium
with a uniform dielectric constant $\epsilon$ and is composed of point
charges. The charge density $Q({\bf r})$ in the system is given by
\begin{eqnarray}
Q({\bf r}) &=& \sum_{\alpha,k} 
q_\alpha \delta({\bf r}-{\bf r}_{\alpha,k})
+ \Sigma({\bf r})
\end{eqnarray}
where $r_{\alpha,k}$ is the position of the $k$th particle of type
$\alpha$, and $\Sigma({\bf r})$ is an applied charge density.

The total electrostatic energy $E$ of the system is given by
\cite{Jackson_1975}
\begin{eqnarray}
E &=& \frac{\epsilon}{8\pi} 
\int d{\bf r} \nabla\phi({\bf r})\cdot\nabla\phi({\bf r})
\end{eqnarray}
where $\phi$ is the electrostatic potential, which can be obtained by
solving Poisson's equation
\begin{eqnarray}
- \epsilon\nabla^2\phi({\bf r}) &=& 4\pi Q({\bf r})
\label{Poisson:eq}
\end{eqnarray}
where $\phi$ is the electrostatic potential, and $Q({\bf r})$ is
the charge density.

The associated Green's function to this problem is:
\begin{eqnarray}
-\frac{\epsilon}{4\pi}
\nabla^2 G_0({\bf r},{\bf r}') &=& \delta({\bf r}-{\bf r}')
\label{G-Poisson:eq}
\end{eqnarray}
where $G_0({\bf r},{\bf r}')$ is the Green's function.  Given the Green's
function, the electrostatic potential can be determine directly from
the charge density distribution $Q({\bf r})$ in the system
\cite{Jackson_1975}:
\begin{eqnarray}
\phi({\bf r}) &=& \int d{\bf r}' G_0({\bf r}',{\bf r}) Q({\bf r}')
\nonumber
\\ && \quad
+ \frac{\epsilon}{4\pi}\oint d{\bf r}' [
G_0({\bf r}',{\bf r})\nabla'\phi({\bf r}')
\nonumber
\\ && \quad
-\phi({\bf r}')\nabla'G_0({\bf r}',{\bf r})]
\label{phi1:eq}
\end{eqnarray}
where the surface integral is taken over the boundary of the system,
with the vector $d{\bf r}'$ pointing outwards from the system, normal
to its surface.  The second term depends only on the boundary
conditions of the system and not on the charge distribution within the
system.

For the problem of an electrolyte confined between two metal plates,
we have Dirichlet boundary conditions and the potential $\phi$ is zero
on the metal surface.  In this situation, we choose $G_0({\bf r}',{\bf
r})=0$ for ${\bf r}'$ on the surface, and so the surface terms in
Eq.~(\ref{phi1:eq}) vanish to yield:
\begin{eqnarray}
\phi({\bf r}) &=& \int d{\bf r}' G_0({\bf r},{\bf r}') Q({\bf r}')
\label{phi2:eq}
\end{eqnarray}

The energy of the electrostatic field can then be written entirely in
terms of the charge distribution
\begin{eqnarray}
E &=& \frac{1}{2}\int d{\bf r} d{\bf r}' 
Q({\bf r}) G_0({\bf r},{\bf r}') Q({\bf r}')
\end{eqnarray}
The expression for the energy given above contains contribution from
the interaction of each ion with itself --- the self-energy.
Subtracting this term from the energy, we write
\begin{eqnarray}
E &=& \frac{1}{2}\int d{\bf r} dr' 
Q({\bf r}) G_0({\bf r},{\bf r}') Q({\bf r}')
\nonumber
\\ && \qquad
- \frac{1}{2}\sum_{\alpha,k} q_{\alpha}^2 
  G_0({\bf r}_{\alpha,k},{\bf r}_{\alpha,k})
\end{eqnarray}

The grand partition function $Z_G$ for a system of point charges at
fixed chemical potential $\mu_\alpha$ is given by
\begin{eqnarray}
Z_G[\gamma] &=& \sum_{N_1=0}^\infty \cdots \sum_{N_M=0}^\infty
\prod_\nu
\frac{1}{N_\nu! \Lambda_\nu^{3N_\nu}}
\nonumber \\ 
&& \qquad \times
\int \prod_{\tau t} d{\bf r}_{\tau t} 
e^{-\beta E+\sum_{\alpha,k}\gamma_\alpha({\bf r}_{\alpha,k})}
\nonumber
\\
&=& \sum_{N_1=0}^\infty \cdots \sum_{N_M=0}^\infty
\prod_\nu
\frac{1}{N_\nu! \Lambda_\nu^{3N_\nu}}
\int \prod_{\tau t} d{\bf r}_{\tau t} 
\nonumber \\ 
&& \qquad \times
  e^{-\frac{\beta}{2}\int d{\bf r}d{\bf r}' 
  Q({\bf r})G_0({\bf r},{\bf r}')Q({\bf r}')}
\nonumber 
\\ && \qquad
\times  
e^{\sum_{\alpha,k} [\gamma_\alpha({\bf r}_{\alpha,k})
  +\frac{\beta}{2}q_{\alpha}^2
    G_0({\bf r}_{\alpha,k},{\bf r}_{\alpha,k})]}
\label{grand_part:eq}
\end{eqnarray}
where $\beta=1/(k_BT)$, $k_B$ is the Boltzmann constant, $N_\nu$ is
the number of ions of type $\nu$, ${\bf r}_{\alpha j}$ is the position
of the $j$th ion of type $\alpha$, $\Lambda_\nu$ is the thermal
wavelength of an ion of type $\nu$, $\gamma_\alpha({\bf
r})=\beta[\mu_\alpha+u_\alpha({\bf r})]$, and $u_\alpha$ is an
external potential acting on ions of type $\alpha$.

Introducing the Hubbard-Stratonovich transformation, the grand
partition function can be expressed in terms of a functional integral
\cite{Kholodenko_Beyerlein_1986,Lue_etal_1999,Ortner_1999,Netz_Orland_2000b}
\begin{eqnarray}
Z_G[\gamma] &=& 
\frac{1}{{\mathcal N}_0}
\int {\mathcal D}\psi(\cdot)
\nonumber
\\ && \quad
\times \exp\Bigg[
- \frac{1}{2\beta}\int d{\bf r}d{\bf r}' 
  \psi({\bf r})G_0^{-1}({\bf r},{\bf r}')\psi({\bf r}')
\nonumber 
\\ && \quad
+ \sum_\alpha \Lambda_\alpha^{-d} 
  \int d{\bf r} e^{\gamma_\alpha({\bf r}) - q_\alpha i\psi({\bf r})}
\nonumber
\\ && \quad
\times e^{\frac{1}{2}\beta q_\alpha^2G_0({\bf r},{\bf r})}
\Bigg]
\label{ZG-form1:eq}
\end{eqnarray}
where
\begin{eqnarray}
{\mathcal N}_0 &=& \int{\mathcal D}\psi(\cdot)
\exp\left[
- \frac{1}{2\beta}\int d{\bf r}d{\bf r}' 
  \psi({\bf r})G_0^{-1}({\bf r},{\bf r}')\psi({\bf r}')
\right].
\end{eqnarray}
The function $i\psi({\bf r})$ can be interpreted as being equal to an
instantaneous value of $\beta\phi({\bf r})$, and the functional
integral can be thought of as an integral over all possible ``shapes''
of the electric field due to the thermal motion of the electrolytes.

\subsection{Mean-field approximation}

In the mean-field approximation, the value of the functional integral
is replaced by the value of the integrand at its saddle point. The
value of the field at the saddle point, denoted by $\bar{\psi}$, is
given implicitly by
\begin{eqnarray}
\frac{\delta\ln Z_G[i\bar{\psi}]}{\delta i\psi(r)} &=& 0,
\nonumber
\end{eqnarray}
which leads to the relation
\begin{eqnarray}
-\frac{\epsilon}{4\pi\beta}\nabla^2 i\bar{\psi}({\bf r}) &=&
\sum_\alpha q_\alpha \Lambda_\alpha^{-d} 
e^{\gamma_\alpha({\bf r})-q_\alpha i\bar{\psi}({\bf r})}
\nonumber
\\ && \qquad
+\Sigma({\bf r}).
\end{eqnarray}
This is precisely the Poisson-Boltzmann equation.  The electric
potential is given by $\phi({\bf r})=k_BTi\bar{\psi}({\bf r})$.

Within the mean-field approximation, the grand potential is given by
\begin{eqnarray}
\ln\bar{Z}_G[\gamma] &=& 
\frac{1}{2\beta}\int d{\bf r}d{\bf r}' 
  i\bar{\psi}({\bf r})G_0^{-1}({\bf r},{\bf r}')i\bar{\psi}({\bf r}')
\nonumber 
\\ && \quad
+ \sum_\alpha \Lambda_\alpha^{-d} 
  \int d{\bf r} e^{\gamma_\alpha({\bf r}) - q_\alpha i\bar{\psi}({\bf r})}
\end{eqnarray}
The density distribution of ion of species $\alpha$ can be determined
by taking the functional derivative of the grand partition function
with respect to $\gamma_\alpha$ \cite{Lebowitz_Percus_1963}, which
yields
\begin{eqnarray}
\bar{\rho}_\alpha({\bf r}) 
&=& \frac{\delta\ln\bar{Z}_G[\gamma] }
  {\delta\gamma_{\alpha}({\bf r})}
\nonumber
\\
&=&\Lambda_\alpha^{-d} 
e^{\gamma_\alpha({\bf r})-q_\alpha i\bar{\psi}({\bf r})}
\end{eqnarray}

\subsection{One-loop expansion}
The functional integral for the grand partition function can be cast
into a form that allows one to perform a loop-wise expansion around
the mean-field solution $\bar{\psi}({\bf r})$.  Defining
$\delta\psi=\psi-\bar{\psi}$ as the deviation from the mean-field
solution, Eq.~(\ref{ZG-form1:eq}) is re-written as a functional
integral over $\delta\psi({\bf r})$. The quadratic term in the action
is now defined by a {\em screened} propagator (Green' s function)
defined by a Dyson equation
\begin{eqnarray}
G^{-1}({\bf r},{\bf r}') 
&=& G_0^{-1}({\bf r},{\bf r}') + \delta({\bf r}-{\bf r}')\bar{K}_2({\bf r}'),
\end{eqnarray}
where the $n$-th moments of the mean-field charge densities have been
defined as
\begin{eqnarray}
\bar{K}_n({\bf r}) &\equiv& 
\beta \sum_\alpha q_\alpha^n \bar{\rho}_\alpha({\bf r}).
\end{eqnarray}
The second moment $\bar{K}_2({\bf r})$ corresponds to a (in general
${\bf r}$-dependent) inverse screening length. The third moment
$\bar{K}_3({\bf r})$ will turn out to be crucial for the physics of
asymmetric electrolytes.  Expanding to order one-loop,
\begin{eqnarray}\label{ZG1}
\ln Z_G^{(1)}[\gamma] &=& 
\ln\bar{Z}_G[\gamma] \\
&-& \frac{1}{2} {\rm Tr} [\ln (1+\bar{K}_2G_0) - \bar{K}_2G_0]\nonumber.
\end{eqnarray}
Here, the last term $\frac{1}{2}{\rm Tr}[\bar{K}_2G_0]$ corresponds to
the self-energy of the charges which has to be subtracted from the
total electrostatic energy.  Equation~(\ref{ZG1}), which is still
completely general, forms the starting point for the discussion of
fluctuation-induced effects in electrolytes.  Since in the following,
all the calculations are to one-loop order, the index $(1)$ in
$Z_G^{(1)}$ is omitted.

\section{Conducting parallel plates}
\label{examples:sec}

%
%

Consider a medium of dielectric constant $\epsilon$ which is
confined between two conducting metal plates separated by a distance
$L$.  In this geometry, the mean-field solution for the field
$\bar{\psi}({\bf r})$ and the densities $\bar{\rho}_\alpha ({\bf r})$
is trivial,
\begin{eqnarray}
i\bar{\psi}({\bf r}) &=& 0,\quad
\bar{\rho}_\alpha ({\bf r}) = \bar{\rho}_\alpha
=\Lambda_\alpha^{-d} e^{\gamma_\alpha}.
\end{eqnarray}
That is, the electric potential between the plates is zero, and the
electrolyte density is uniform and solely determined by the chemical
potentials $\mu_\alpha$ and the temperature $k_BT=1/\beta$
($\gamma_\alpha=\beta\mu_\alpha$).

The (screened) Green's function $G({\bf r},{\bf r}')$ for the
Poisson-Boltzmann equation is obtained from a Fourier series for the
screened Coulomb interaction and given by
\begin{widetext}
\begin{eqnarray}
G({\bf r},{\bf r}') &=& \frac{4\pi}{\epsilon} \frac{2}{L}
\sum_{n=1}^\infty \int \frac{dq_1}{2\pi} \frac{dq_2}{2\pi}
\left[\left(\frac{n\pi}{L}\right)^{2}+q_1^2+q_2^2+\kappa^2\right]^{-1}
\sin\left(\frac{n\pi z}{L}\right) \sin\left(\frac{n\pi z'}{L}\right)
e^{iq_1(x-x')+iq_2(y-y')},
\label{G:eq}
\end{eqnarray}
\end{widetext}
where $\kappa$ is the inverse Debye screening length in the mean-field
approximation, which is defined as
\begin{eqnarray}
\kappa^2 &\equiv& \frac{4\pi}{\epsilon}\bar{K}_2=
\frac{4\pi\beta}{\epsilon} 
\sum_\alpha q_\alpha^2 \bar{\rho}_\alpha
\end{eqnarray}
The (unscreened) Green's function $G_0({\bf r},{\bf r}')$ of the {\em
Poisson} equation (see Eqs.~(\ref{Poisson:eq}) and
(\ref{G-Poisson:eq})), is obtained from Eq.~(\ref{G:eq}) by setting
$\kappa=0$.

\subsection{Density profile}
All the thermodynamic properties of the system can be derived from the
grand potential.  The density profile to one loop order is given by
the functional derivative \cite{Lebowitz_Percus_1963}
\begin{eqnarray}
\rho_\alpha({\bf r}) &=& \frac{\delta \ln Z_G}{\delta\gamma_\alpha({\bf r})}
\approx \bar{\rho}_\alpha \left[ 
1 - \frac{\beta q_\alpha^2}{2} \Delta G({\bf r},{\bf r})
\right.
\\ && \qquad
\left.
+ \frac{\beta q_\alpha}{2} \int d{\bf r}' \Delta G({\bf r}',{\bf r}')
\bar{K}_3({\bf r}')G({\bf r}',{\bf r}) 
\right],\nonumber
\label{density:eq}
\end{eqnarray}
where %
\begin{eqnarray}
\Delta G({\bf r},{\bf r}) &\equiv& G({\bf r},{\bf r})-G_0({\bf r},{\bf r}) 
= -\frac{2}{\epsilon L} I_1(\kappa L,z)
\end{eqnarray}
with $I_1$ defined as
\begin{eqnarray}\label{I1definition}
I_1(\kappa L,z) &\equiv& 
\sum_{n=1}^\infty
\sin^2\left(\frac{n\pi z}{L}\right)
\ln\left(1+ \left(\frac{\kappa L}{n\pi}\right)^2\right)
\end{eqnarray}
Note that in the bulk limit where $L\to\infty$, $G({\bf r},{\bf
r}')=e^{-\kappa|{\bf r}-{\bf r}'|}/(\epsilon|{\bf r}-{\bf r}'|)$, and
$\Delta G(r,r)=-\kappa/\epsilon$.  In this case, the density profile
becomes uniform and is given by
\begin{eqnarray}
\rho_\alpha^{\rm bulk}
&\approx& \Lambda_\alpha^{-d} e^{\gamma_\alpha} \left[ 1
+ \frac{\beta q_\alpha^2\kappa}{2\epsilon}
- \frac{4\pi\beta q_\alpha}{2\epsilon\kappa} \bar{K}_3
\right] 
\end{eqnarray}

For {\em symmetric electrolytes}, where the magnitude of all the
charges of the particles is the same (i.e., $|q_\alpha|=q$ for all
$\alpha$, and therefore $\bar{K}_3=0$), the last term in
Eq.~(\ref{density:eq}) vanishes, and the expression for the density
profile reduces to
\begin{eqnarray}
\rho_\alpha({\bf r}) 
&\approx& \Lambda_\alpha^{-d} e^{\gamma_\alpha} \left[ 1
+ \frac{\beta q_\alpha^2}{\epsilon L} I_1(\kappa L,z)
\right] 
\label{density-sym:eq}
\end{eqnarray}
The second term is the only correction term for a symmetric
electrolyte, and its sign is independent of the sign of the charge.
As a result, the fluctuation corrections increase the electrolyte
concentration of the system, as compared to the mean-field
approximation.  However, the ion concentration at the metal surface
remains at its mean-field value.  In addition, the system is
everywhere locally neutral.

\begin{figure}
\resizebox*{\columnwidth}{!}{
\includegraphics[clip]{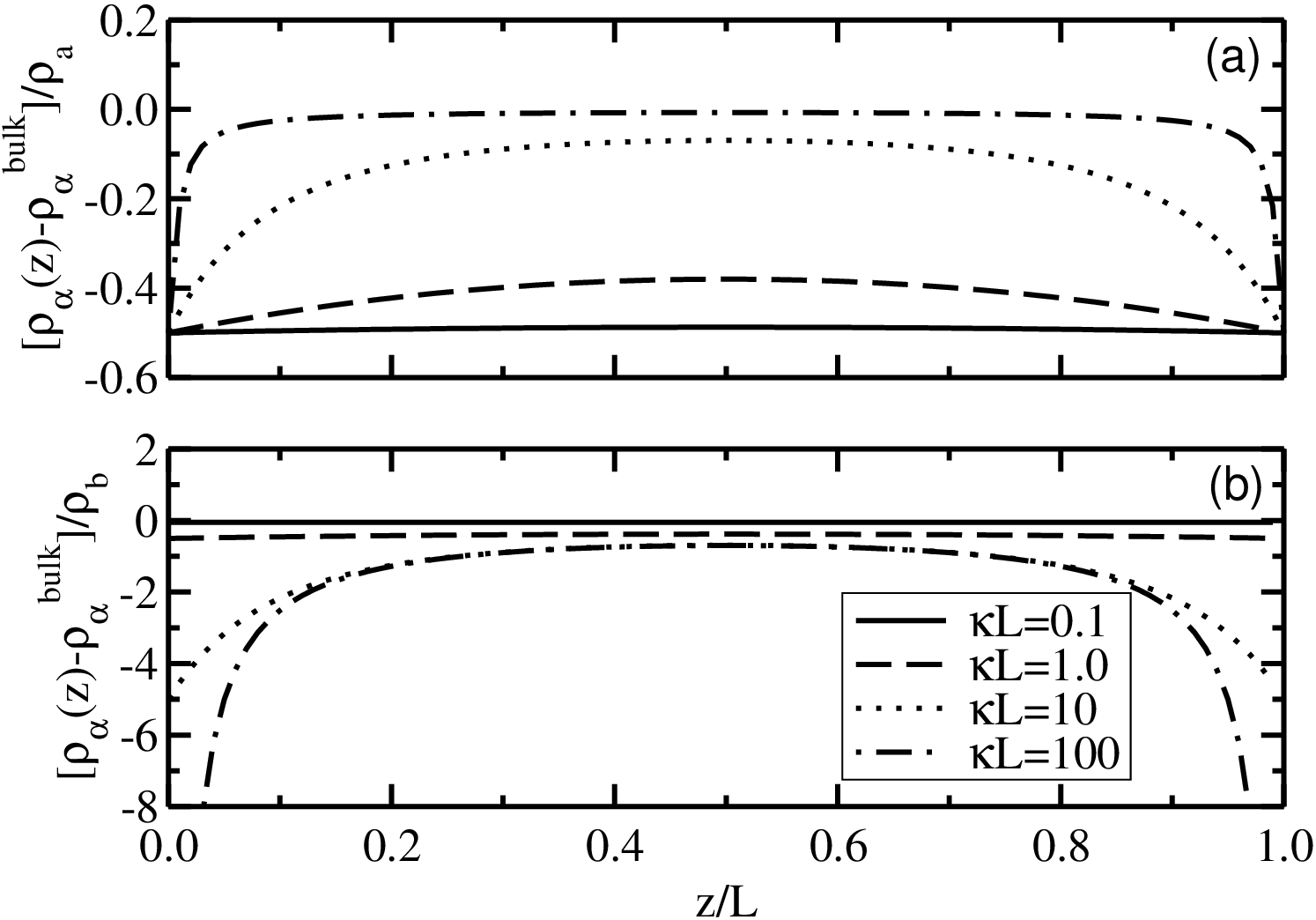}
}
\caption{\label{density1-symmetric:fig}
Density profile for a symmetric electrolyte solution confined
between two conducting plates with (a) $\kappa$ held constant
($\rho_a=\beta\bar{\rho}_\alpha q_\alpha^2\kappa/\epsilon$)
and with
(b) $L$ held constant 
($\rho_b=\beta\bar{\rho}_\alpha q_\alpha^2/(\epsilon L)$): 
(i) $\kappa L=0.1$ (solid line),
(ii) $\kappa L=1$ (dashed line),
(iii) $\kappa L=10$ (dotted line), and
(iv) $\kappa L=100$ (dashed-dotted line).
}
\end{figure}

The difference between the density between the plates and that of a
bulk system at the same chemical potential is given by
\begin{eqnarray}
\rho_\alpha({\bf r})-\rho_\alpha^{\rm bulk}
&\approx&  
\frac{\beta \bar{\rho}_\alpha q_\alpha^2}{\epsilon L} \left[
I_1(\kappa L,z) - \frac{\kappa L}{2}
\right] 
\end{eqnarray}
Plots of the density profile difference between the confined system
and the bulk system at different values of $\kappa L$ are given in
Fig.~\ref{density1-symmetric:fig}.  In
Fig.~\ref{density1-symmetric:fig}a, the plate separation $L$ is
varied while holding the value of $\kappa$ (i.e., the ion chemical
potentials) fixed; in Fig.~\ref{density1-symmetric:fig}b, the value
of $\kappa$ is varied while the plate separation is held fixed.

The ions are more concentrated in the center of the system than near
the plates.  At the center of the plates, where the ion concentration
is at a maximum, the density can be computed in closed form, to yield:
\begin{eqnarray}
\rho_\alpha\left(\frac{L}{2}\right)
&=& \bar{\rho}_\alpha \left[
1 + \frac{\beta q_\alpha^2}{\epsilon L}
  \ln\cosh\frac{\kappa L}{2}
\right]
\end{eqnarray}
In addition, for a symmetric electrolyte, the concentration of the
ions in the confined system is lower than that of the bulk system at
the same chemical potentials.

The local {charge distribution} $Q({\bf r})$ between the plates can be
computed from the density distribution of the ions:
\begin{eqnarray}\label{chargedensity}
Q({\bf r}) &\equiv& \sum_\alpha q_\alpha \rho_\alpha({\bf r})
\\
&=& - \frac{\bar{K}_3}{2}  \left\{\Delta G({\bf r},{\bf r})
- \bar{K}_2 \int d{\bf r}'\Delta G({\bf r}',{\bf r}')G({\bf r}',{\bf r})
\right\}
\nonumber
\\
&=& \frac{2\sigma^*}{L}  \Bigg\{
I_1(\kappa L,z)- (\kappa L)^2 I_2(\kappa L,z)
\nonumber
\\ && \qquad
+ \frac{\sinh\frac{\kappa z}{2}
  \sinh\frac{\kappa L}{2}(z/L-1)}{\cosh\frac{\kappa L}{2}}
  (\kappa L)^2 I_3(\kappa L)
\Bigg\},\nonumber
\end{eqnarray}
where we have used the fact that $\sum_\alpha
q_\alpha\bar{\rho}_\alpha=0$, defined
$\sigma^*\equiv\bar{K}_3/(2\epsilon)$, and introduced the functions
(cf.\ Appendix \ref{appintegrals})
\begin{eqnarray}\label{I23definition}
I_2(\kappa L,z) &=& 
  \sum_{m=1}^\infty
  \frac{\ln\left(1+\left(\frac{\kappa L}{m\pi}\right)^2\right)}
       {(2m\pi)^2+(\kappa L)^2}
\sin^2\frac{m\pi z}{L}\\
I_3(\kappa L) &=&
\frac{1}{(\kappa L)^2} \sum_{m=1}^\infty
\frac{(2m\pi)^2 \ln\left(1+\left(\frac{\kappa L}{m\pi}\right)^2\right)}
     {(2m\pi)^2+(\kappa L)^2}.
\nonumber
\end{eqnarray}
In the case where $\bar{K}_3=0$ (e.g.,
a symmetric electrolyte with $|q_\alpha|=q$ for all $\alpha$), 
$Q({\bf r})$ is trivially zero, and the system 
(apart from being locally neutral everywhere) is globally neutral.

A non-trivial situation arises for asymmetric electrolytes with
$\bar{K}_3\ne0$. As shown below, the fluctuations around the
mean-field solution lead to a non-zero charge distribution
between the metal plates with a density profile that strongly depends
on the dimensionless parameter $\kappa L$.

\begin{figure}
\resizebox*{\columnwidth}{!}{
\includegraphics[clip]{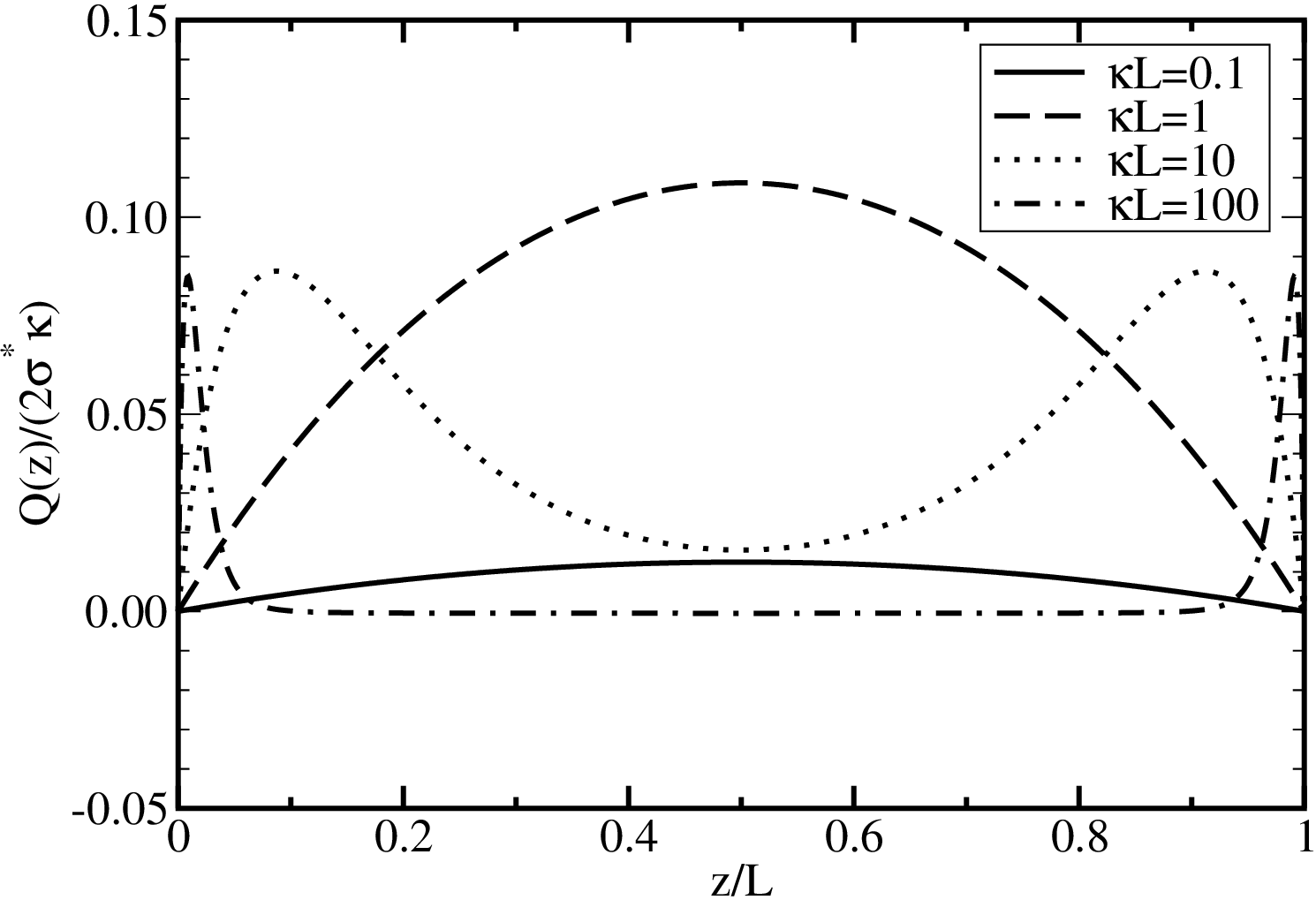}
}
\caption{\label{charge:fig}
Variation of the charge density for an electrolyte solution confined
between two conducting plates at constant $\kappa$: 
(i) $\kappa L=0.1$ (solid line),
(ii) $\kappa L=1$ (dashed line), 
(iii) $\kappa L=10$ (dotted line), and
(iv) $\kappa L=100$ (dashed-dotted line).
}
\end{figure}

In Fig.~\ref{charge:fig}, the variation of the local charge
distribution for different values of $\kappa L$ is plotted at constant
$\kappa$.  For low values of $\kappa L$, the local charge distribution
has a single peak located in the center of the system with the same
sign as $\bar{K}_3$.  For a binary asymmetric electrolyte, this
implies that the sign of the net local charge is the same as that of
the ion with the larger charge magnitude.

As $\kappa L$ further increases, the charge distribution initially
uniformly increases; however, at a critical value of $\kappa L$, the
single peak becomes a local minimum, and two peaks emerge on either
side of this minimum.  These peaks move towards the boundaries with
increasing $\kappa L$, while the value of the charge density at the
center of the system decreases, eventually changing sign.  This is
similar to the change in the monomer distribution for a
polyelectrolyte confined between two oppositely charged plates
\cite{Podgornik_Dobnikar_2001}.

\begin{figure}
\resizebox*{\columnwidth}{!}{
\includegraphics[clip]{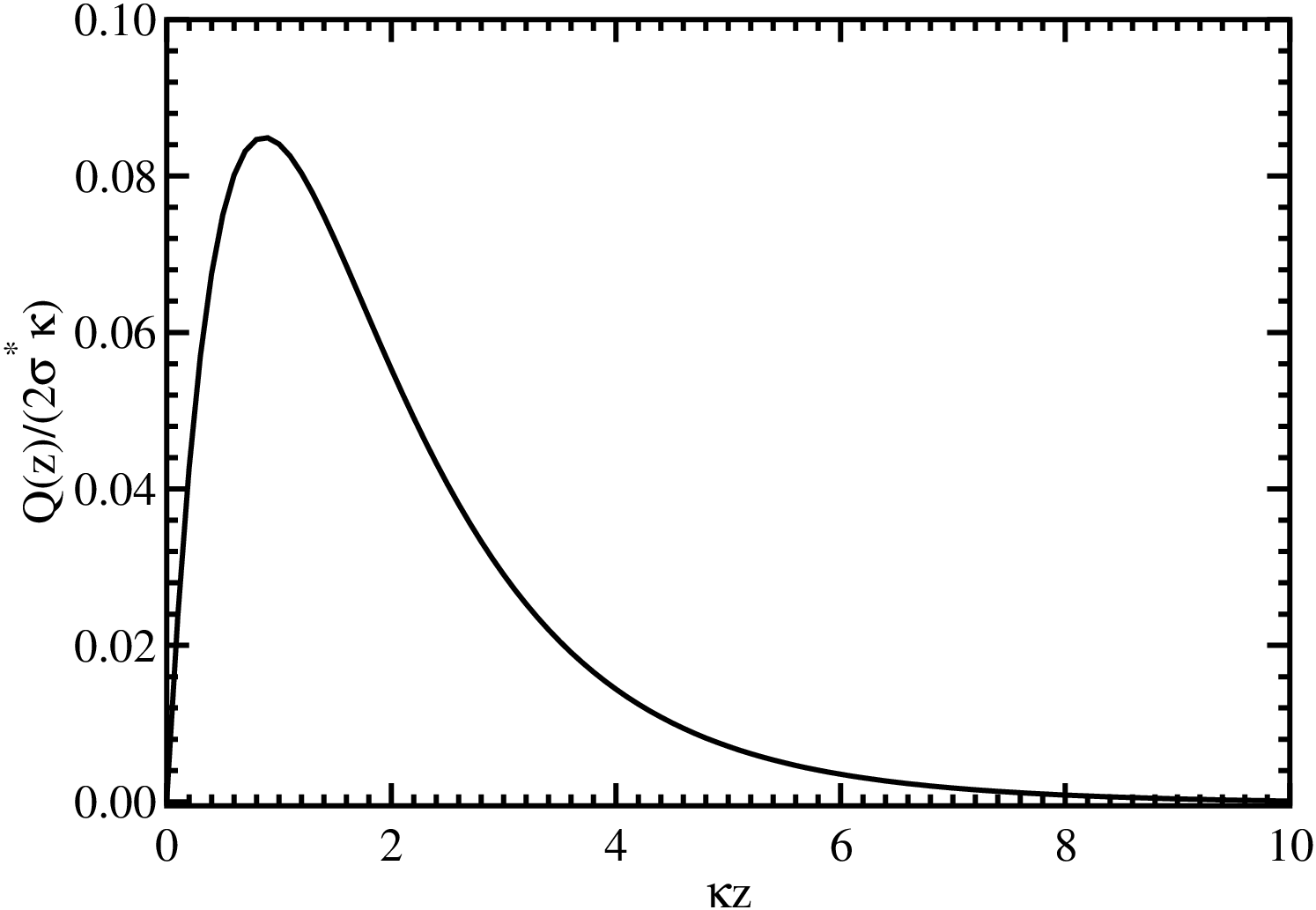}
}
\caption{\label{charge-limit:fig} 
The local charge distribution near a
metal plate in the limit of infinite plate separation.}
\end{figure}

As the plates become further and further separated, the charge density
near the plates approaches a limiting form $Q_{\infty}(z)\equiv
\lim_{\kappa L \to \infty} Q(z)/(2\kappa\sigma^*)$, which can be
written analytically as:
\begin{eqnarray}
Q_{\infty}(z)
&=&\frac{1}{4\kappa z}(e^{-2\kappa z}-1)
+\left(\frac{1}{2}-\frac{\log 3}{8}\right) e^{-\kappa z} 
\nonumber
\\ && \quad
+\frac{e^{\kappa z}}{8}
\left[ \mbox{\rm Ei}(-3\kappa z)-\mbox{\rm Ei}(-\kappa z)\right]
\\
&& \quad 
- \frac{e^{-\kappa z}}{8}
\left[ \mbox{\rm Ei}(-\kappa z)-\mbox{\rm Ei}(\kappa z) \right]\nonumber
\end{eqnarray}
where ${\rm Ei}$ denotes the exponential integral.  This function is plotted
in Fig.~\ref{charge-limit:fig}.

Integrating the distribution of charge contained between the plates,
the total net charge per unit area ${\mathcal Q}$ is determined to be
\begin{eqnarray}
{\mathcal Q} &=& \int_0^L dz Q({\bf r})
= 2\sigma^*(\kappa L)\tanh\frac{\kappa L}{2} I_3(\kappa L).
\end{eqnarray}
Therefore, an asymmetric electrolyte system has a non-vanishing net
charge.  As shown in the next section, this net charge is
counter-balanced by an induced surface charge on the metal plates.

\subsection{Electric potential}

The electric potential $\phi({\bf r})$ between the plates can be obtained
from the grand partition function
\cite{Netz_Orland_2000b}
\begin{eqnarray}
\phi({\bf r})
&=& \frac{1}{\beta}i\psi({\bf r}) 
= \frac{1}{\beta} \frac{\delta\ln Z_G}{\delta\Sigma({\bf r})}
\nonumber
\\
&\approx &
- \frac{\bar{K}_3}{2} 
\int dr' \Delta G(r',r')  G(r',r)
\nonumber
\\
&\approx& 
- \frac{8\pi\sigma^*L}{\epsilon}
\Bigg[ 
\frac{\sinh\frac{\kappa z}{2}
  \sinh\frac{\kappa L}{2}(z/L-1)}{\cosh\frac{\kappa L}{2}}
I_3(\kappa L)
\nonumber
\\ && \qquad
- I_2(\kappa L,z)
\Bigg]
\label{psi:eq}
\end{eqnarray}
In the case of a symmetric electrolyte, the correction term vanishes,
and the electric field is identically zero between the plates, similar
to the mean-field approximation.  In the case of an asymmetric
electrolyte, however, the fluctuation correction is non-zero, and the
electric potential varies within the cavity.

\begin{figure}
\begin{center}
\resizebox*{\columnwidth}{!}{
\includegraphics[clip]{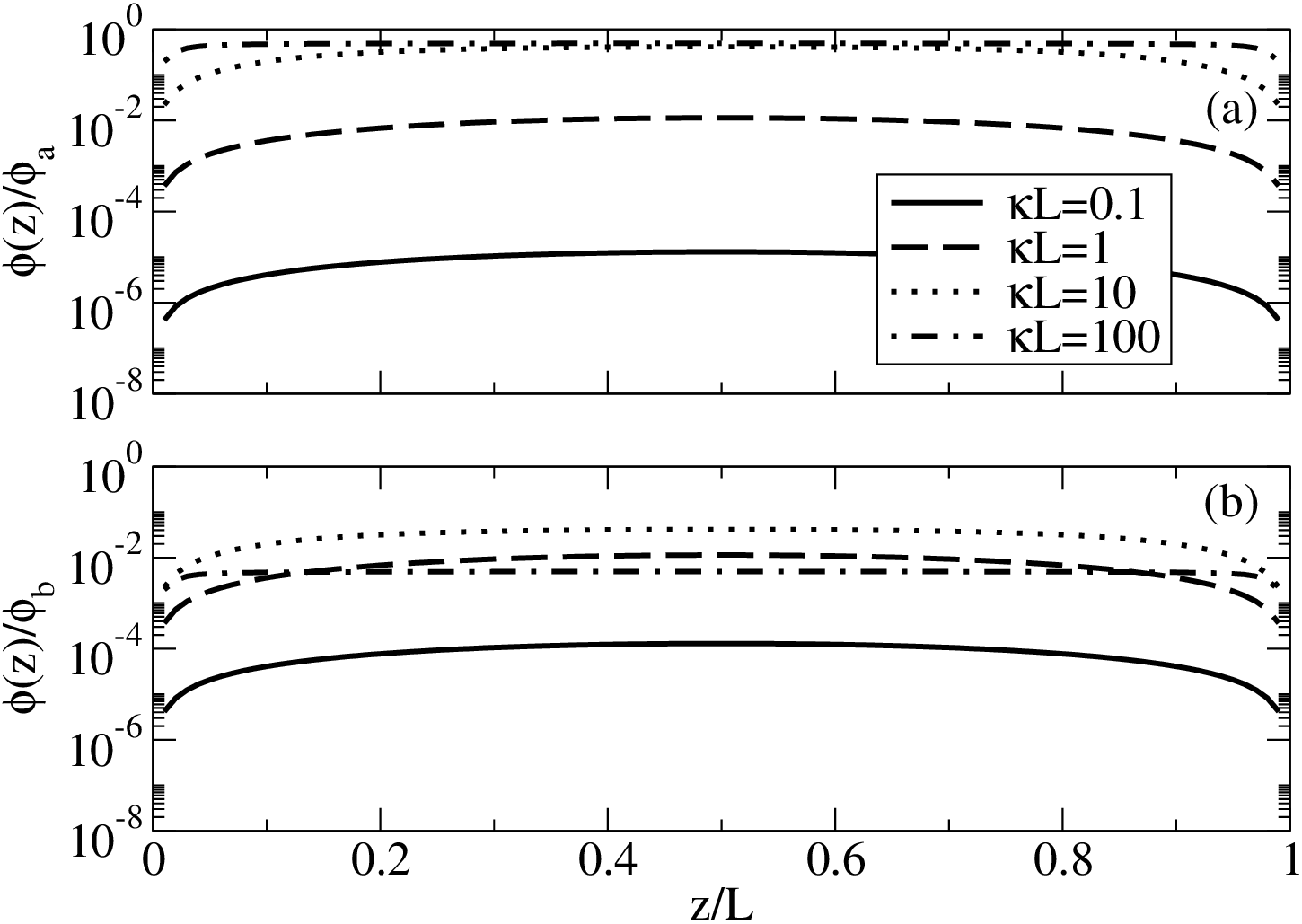}
}
\end{center}
\caption{\label{psi:fig}
Variation of the electric potential between the metal plates
at (a) fixed $\kappa$ ($\phi_a=-8\pi\sigma^*/(\epsilon\kappa)$) and
(b) fixed plate separation $L$  ($\phi_b=-8\pi\sigma^*L/\epsilon$):
(i) $\kappa L=0.1$ (solid line),
(ii) $\kappa L=1$ (dashed line),
(iii) $\kappa L=10$ (dotted line), and
(iv) $\kappa L=100$ (dashed-dotted line).
}
\end{figure}

In Fig.~\ref{psi:fig}, the electric potential $\phi(z)$ in the system
is plotted.  If the plate distance is kept fixed, then the electric
potential between the plates uniformly increases with $\kappa$ to a
constant profile (see Fig.~\ref{psi:fig}a).  If $\kappa$ is held fixed
while the distance between the plates $L$ is increased, then the
potential increases from zero to a maximum value at a critical value
of $L$, and subsequently decrease to zero (see Fig.~\ref{psi:fig}b).

As a result of the different density distribution between ion species,
local charge densities develop.  The electrolyte system is no longer
necessarily electrically neutral overall.  The metal plates, however,
develop a {\em surface charge}, which keeps the overall system globally
neutral.  The magnitude of this surface charge is given by
\begin{eqnarray}
\sigma &=& -\frac{\epsilon}{4\pi}
\left.\frac{\partial\phi({\bf r})}{\partial z}\right|_{z=0}
\nonumber
\\
&\approx& - \sigma^* 
(\kappa L)\tanh\frac{\kappa L}{2} I_3(\kappa L)
\end{eqnarray}
Note that the surface charge density on each of the plates is
precisely half of the net charge of the electrolytes between the
plates (i.e., $\sigma=-{\mathcal Q}/2$), so the system as a whole is
electrically neutral.  

\begin{figure}
\resizebox*{\columnwidth}{!}{
\includegraphics[clip]{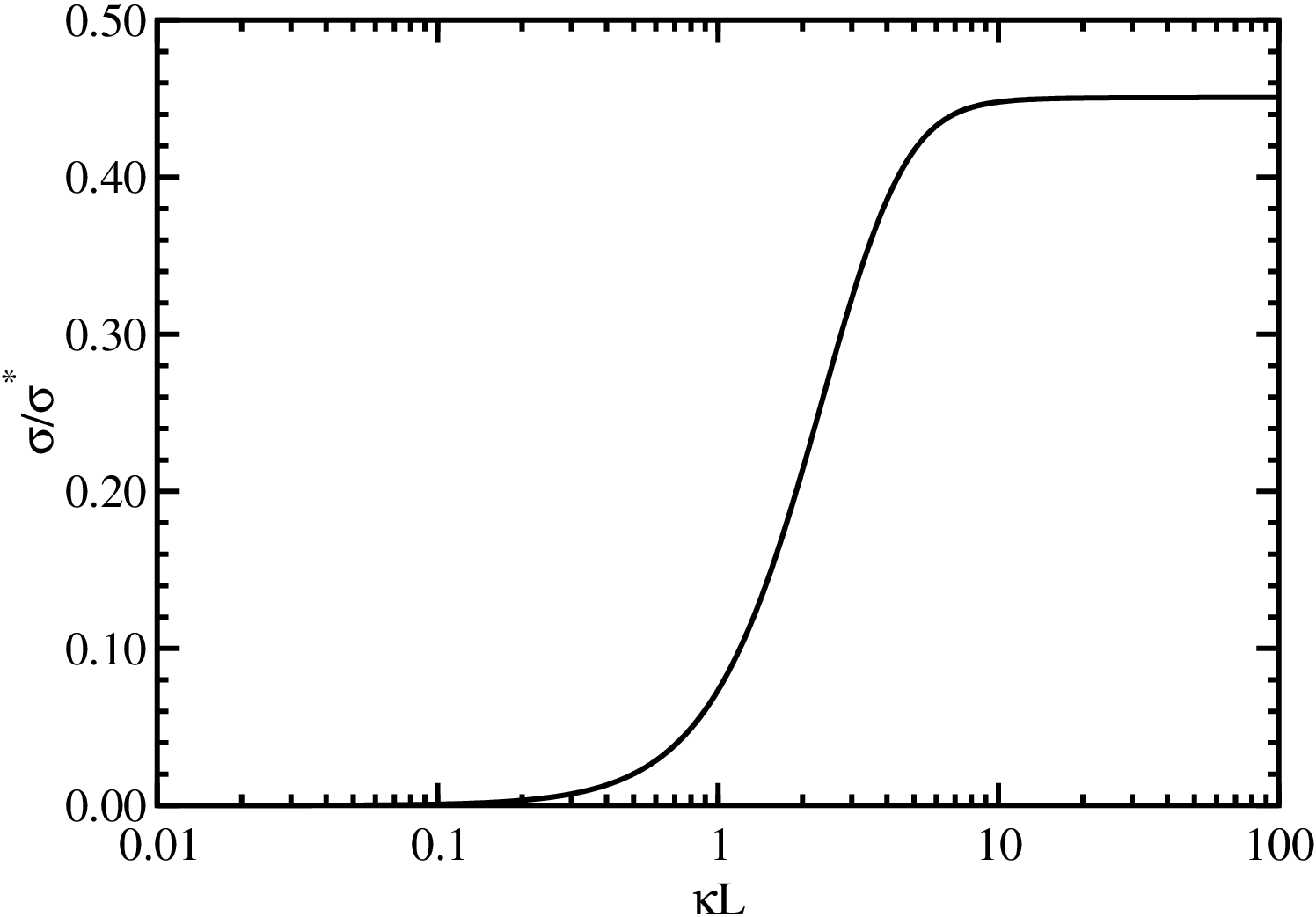}
}
\caption{\label{surf-charge:fig}
Variation of the surface charge density with $\kappa L$.
Note $\sigma^*=-\bar{K}_3/(2\epsilon)$.
}
\end{figure}

In Fig.~\ref{surf-charge:fig}, we plot the variation of the surface
charge on the metal plates as a function of $\kappa L$.  As $\kappa L$
increases, the surface charge increases, reaching a constant non-zero
value $\sigma/\sigma^*=1-\ln 3/2\approx0.451$ as $\kappa L$ approaches
infinity.  This implies that in the limit where the plates are
infinitely far apart, the fluctuations in an asymmetric electrolyte
still induce a surface charge.

\subsection{Casimir Force}
We now consider the situation where two, grounded metal plates are
immersed in a bulk electrolyte solution.  The chemical potentials of
the electrolytes in the bulk solution are fixed at
$\gamma_\alpha=\beta\mu_\alpha$, which are identical to the chemical
potentials of the electrolytes between the plates.  We assume that the
thickness of the metal plates is much larger than the size of the
electrolytes and much larger than the separation between the plates.

In order to obtain the force between the two plates, we calculate the
pressure which is given by
\begin{eqnarray}\label{pdefinition}
p &=& k_BT\frac{\partial \ln Z_G}{\partial V}
= \frac{k_BT}{A}\frac{\partial \ln Z_G}{\partial L}
\end{eqnarray}
where $A$ is the surface area of the plates, and $V=AL$ is the total
volume of the system. We therefore need the full expression for the
grand potential $Z_G$, which to order one-loop is explicitely obtained
from Eq.~(\ref{ZG1}) and can be written as (see Appendix
\ref{ZGappendix})
\begin{eqnarray}\label{ZGresult1}
\ln Z_G[\gamma] &=& 
AL \sum_\alpha\bar{\rho}_\alpha
- \frac{\kappa^2 A}{8\pi} I(\kappa L)
\nonumber
\\
I(\kappa L) &\equiv& -\frac{2}{(\kappa L)^2}\int_0^{\kappa L}dx 
x\ln\left(\frac{\sinh x } {x}\right).
\end{eqnarray}
For large values of $L$, the the grand potential behaves
as
\begin{eqnarray}\label{largeLexp}
\ln Z_G[\gamma] &=& AL \sum_\alpha\bar{\rho}_\alpha
+ AL \frac{\kappa^3}{12\pi} - A
\left(\ln2-\frac{1}{2}\right)\frac{\kappa^2}{8\pi} 
\nonumber
\\ && \qquad
- A\frac{\kappa^2}{8\pi} \ln(\kappa L)
- \frac{A}{L^2} \frac{\zeta(3)}{16\pi} 
+\cdots
\end{eqnarray} 
where the higher order terms vanish more rapidly than $L^{-2}$ with
increasing $L$.

The first two terms are proportional to the volume of the system.
This is essentially the Debye-H\"uckel approximation (order one-loop)
for the fluctuation correction for a bulk electrolyte.  The third term
is proportional to the surface area of the metal plates and is
independent of the separation distance $L$.  This term is esentially
the free energy of creating the metal/electrolyte interfaces.  The
higher order terms represent the free energy cost of separating the
plates.  These terms are responsible for inducing a net attraction
between the metal plates, as demonstrated next.

Owing to Eqs.~(\ref{pdefinition}) and (\ref{ZGresult1}), the pressure
in between the metal plates, due to the presence of the electrolytes,
is given by
\begin{eqnarray}
\frac{p(L)}{k_BT} &=&
\sum_\alpha \bar{\rho}_\alpha
- \frac{\kappa^3}{8\pi} I'(\kappa L)
\end{eqnarray}
Using the large-$L$ expansion  Eq.~(\ref{largeLexp}),
for $\kappa L\gg 1$ the pressure behaves as
\begin{eqnarray}
\frac{p(L)}{k_BT} &=&
\sum_\alpha \bar{\rho}_\alpha
+ \frac{\kappa^3}{12\pi}
- \frac{\kappa^2}{8\pi}\frac{1}{L}
+  \frac{\zeta(3)}{8\pi} \frac{1}{L^3}
+ \cdots
\end{eqnarray}
The sum of the first two terms on the right-hand side are equal to the
pressure of the bulk solution (i.e., the solution outside the plates).
This pressure acts on the outside surface of the plates.

\begin{figure}
\resizebox*{\columnwidth}{!}{
\includegraphics[clip]{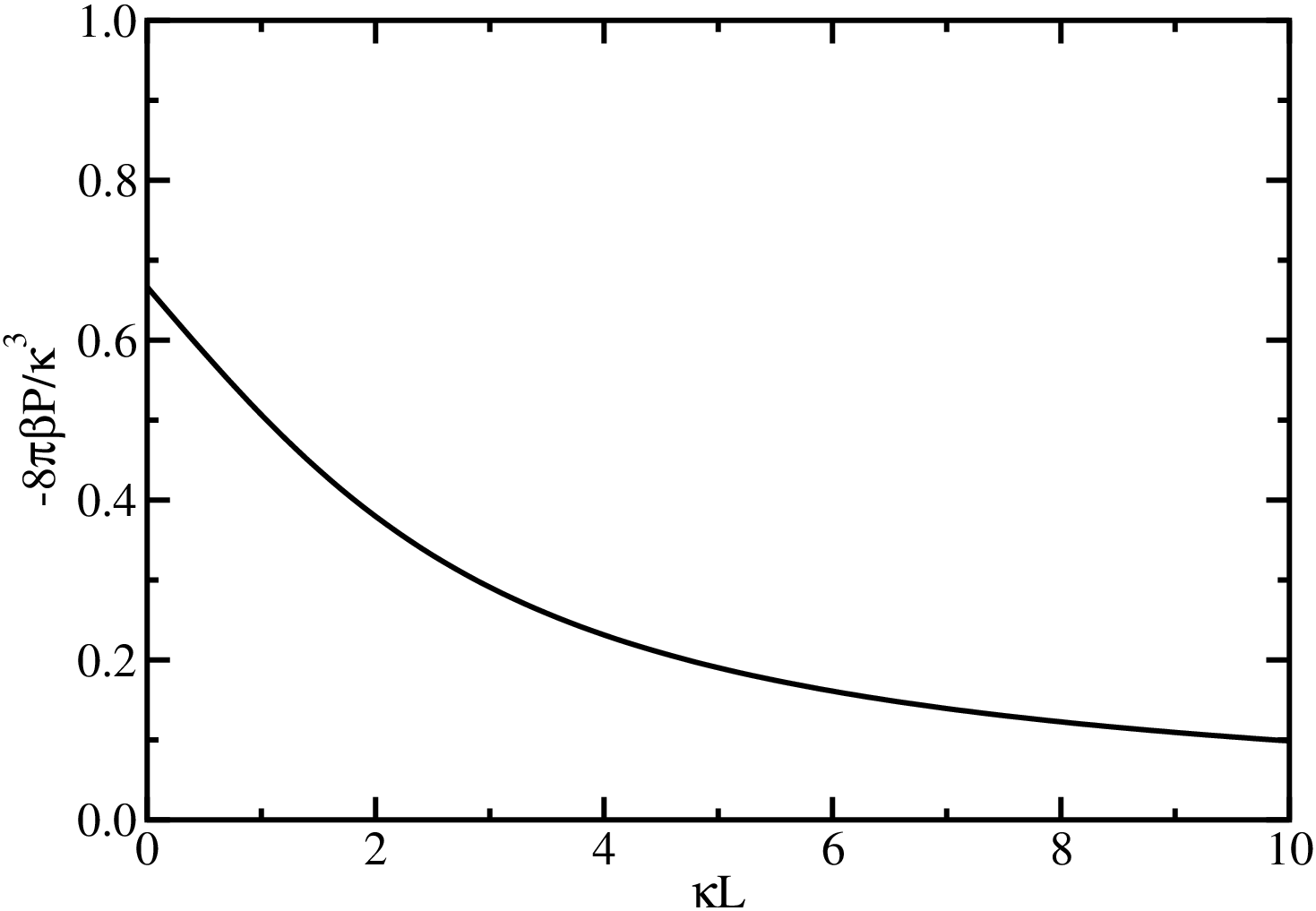}
}
\caption{\label{force:fig} 
Fluctuation induced attraction between two parallel metal plates
separated by a distance $L$.  }
\end{figure}

The {\em net attractive force} per unit area ${\mathcal P}$ acting between
the plates is given by the difference in the pressures inside and
outside the plates:
\begin{eqnarray}
{\mathcal P}(L) &=& p(L,\gamma)-p(\infty,\gamma)
\nonumber
\\
&=& - \frac{k_BT\kappa^3}{8\pi}\left[
I'(\kappa L) + \frac{2}{3}
\right]
\end{eqnarray}
In Fig.~\ref{force:fig}, the variation of this attractive
force with plate separation is plotted, while holding the inverse screening
length $\kappa$ constant.

For small values of $\kappa L$ (e.g., at small plate separations), the
force between the plates behaves as
\begin{eqnarray}
{\mathcal P}(L) 
&\approx&
- \frac{k_BT\kappa^3}{12\pi}\left[
1 - \frac{\kappa L}{4} + \frac{(\kappa L)^3}{90}
+ \cdots
\right]
\end{eqnarray}
The leading term is proportional to the square root of the
temperature.  In the limit that the plate separation vanishes, we see
the force approaches its maximum value, which is finite.

For large values of $\kappa L$ (e.g., for large plate separations),
the force behaves as
\begin{eqnarray}
{\mathcal P}(L) 
&\approx&
- \frac{k_BT}{8\pi}
\frac{\kappa^2}{L} \left[
1 - \frac{\zeta(3)}{(\kappa L)^2}
+ \cdots 
\right]
\end{eqnarray}
The leading order term is independent of temperature.  For large plate
separations, the force decays inversely with the plate separation.


\section{Conclusions}
\label{conclusions:sec}

We have performed order one-loop calculations for an electrolyte
confined between a pair of grounded, metal plates.  For this system,
we find that the fluctuation corrections lead to qualitatively
different physics than predicted by the mean-field approximation.  

The density profiles of the ions are non-uniform.  For symmetric
electrolytes, the densities of the ions have their minimum values
(give by the mean-field values) at the metal interface and are peaked
in the center of the system.  The magnitude of the density profiles is
always lower than the densities of the ions in an equilibrium bulk
solution.  Despite the non-uniform density distribution, the system
still obeys local electrical neutrality, and the electric potential is
everywhere zero.

For asymmetric electrolytes, local charge separation can occur.  When
the plate separation is much less than a screening length, the charge
profile has a single peak at the center of the system.  When the plate
separation becomes much greater than the screening length, the charge
profile develops two peaks, with a minimum in the center of the
system.  This non-uniform charge distribution generates a non-zero
electrical potential throughout the system.  In addition, a surface
charge arises on the metal plates.  In the limit of infinite plate
separation, this surface charge approaches a finite limit
$\sigma/\sigma^*\approx0.451$.  In addition, the charge distribution
near the plates approaches a limiting form.

When the plates are immersed in a bath of electrolytes, the ion-ion
correlations induce an attractive force between them.  This force is
greatest when the plates are touching and decays as $1/L$ at large
plate separation.

In these calculations, the size of the ions was neglected.
Consequently, the predictions are valid only in the limit where the
plate separations are much greater than the ion sizes.  Excluded
volume interactions, as well as other types of forces, can be
incorporated into the theory by changing the starting partition
function Eq.~(\ref{grand_part:eq})
\cite{Lue_etal_1999,Burak_Andelman_2000}.  This method can also be
used to study other types of electrolyte systems, such as
polyelectrolytes or even quantum systems.  This work is currently
being pursued.


\begin{appendix}

\section{Integrals for density}\label{appintegrals}
For the evaluation of the charge density profile in the asymmetric
case $\bar{K}_3\ne 0$, Eq.(\ref{chargedensity}), one requires the
following integrals: 
\begin{eqnarray}
&&\int_0^L dz' \Delta G(z',z') \sin\frac{n\pi z'}{L}\\
&=&
\frac{1}{\epsilon}\sum_{m=1}^\infty 
\ln\left(1+\left(\frac{\kappa L}{m\pi}\right)^2
\right)
\frac{(2m\pi)^2(1-\cos n\pi)}{n\pi[(n\pi)^2-(2m\pi)]}\nonumber
\end{eqnarray}
and 
\begin{eqnarray}
& &\int d{\bf r}'\Delta G({\bf r}',{\bf r}') G({\bf r}',{\bf r}) \\
&=& \frac{8\pi}{\epsilon L} \sum_{n=1}^\infty \frac{\sin \frac{n\pi z}{L} }{
    \left(\frac{n\pi}{L}\right)^2+\kappa^2}
\int_0^L dz'  \Delta G(z',z') \sin\frac{n\pi z'}{L} 
\nonumber
\\
&=& \frac{8\pi L}{\epsilon^2} \sum_{m=1}^\infty 
(2m\pi)^2 \ln\left(1+\left(\frac{\kappa L}{m\pi}\right)^2\right)
\nonumber
\\ && \qquad
\times
\sum_{n=1}^\infty 
\frac{(1-\cos n\pi)\sin\frac{n\pi z}{L}}{n\pi[(n\pi)^2+(\kappa L)^2][(n\pi)^2-(2m\pi)]}\nonumber
\end{eqnarray}
To perform the summation over the index $n$, we first note the
following relation,
\begin{eqnarray}
\sum_{n=1}^\infty \frac{n\pi(1-\cos n\pi)\sin\frac{n\pi z}{L}}{(n\pi)^2+a^2}
&=& \frac{\cosh\frac{a}{2}(2z/L-1)}{2\cosh(a/2)}
\end{eqnarray}
which can be verified by expanding the right-side of the expression in
a Fourier sine series.  In addition, we note that
\begin{eqnarray}
&& \frac{1}{(n\pi)[(n\pi)^2+(\kappa L)^2][(n\pi)^2-(2m\pi)^2]} 
\\
&& \quad =\frac{1}{(2m\pi)^2+(\kappa L)^2}
\left\{\frac{1}{(\kappa L)^2}\left[ 
\frac{n\pi}{(n\pi)^2+(\kappa L)^2}-\frac{1}{n\pi}\right]
\right.
\nonumber
\\  
&& \qquad\qquad + \left.\frac{1}{(2m\pi)^2}\left[
\frac{n\pi}{(n\pi)^2-(2m\pi)^2}- \frac{1}{n\pi}\right]
\right\}.\nonumber
\end{eqnarray}
This is used when performing the summation, 
\begin{widetext}
\begin{eqnarray}\label{GGcalc}
\int d{\bf r}'\Delta G({\bf r}',{\bf r}') G({\bf r}',{\bf r}) 
&=& \frac{8\pi L}{\epsilon^2}
\sum_{m=1}^\infty\frac{(2m\pi)^2}{(2m\pi)^2+(\kappa L)^2}
  \ln\left(1+\left(\frac{\kappa L}{m\pi}\right)^2\right)
\frac{1}{2}\left\{
\frac{1}{(\kappa L)^2}
\left[
\frac{\cosh\frac{\kappa L}{2}(2z/L-1)}{\cosh\frac{\kappa L}{2}}-1
\right]
\right.
\nonumber
\\ && \qquad
\left.
+ \frac{1}{(2m\pi)^2}
\left[\frac{\cos m\pi(2z/L-1)}{\cos m\pi}-1
\right]
\right\}
\nonumber
\\
&=& \frac{4\pi L}{\epsilon^2} \frac{1}{(\kappa L)^2}
\sum_{m=1}^\infty\frac{(2m\pi)^2}{(2m\pi)^2+(\kappa L)^2}
  \ln\left(1+\left(\frac{\kappa L}{m\pi}\right)^2\right)
\nonumber
\\ && \qquad
\times \left\{
\left[
\frac{\cosh\frac{\kappa L}{2}(2z/L-1)}{\cosh\frac{\kappa L}{2}}-1
\right]
+ \left(\frac{\kappa L}{2m\pi}\right)^2\left(\cos\frac{2m\pi z}{L}-1\right)
\right\}
\nonumber
\\
&=& \frac{4\pi L}{\epsilon^2} \frac{1}{(\kappa L)^2}
\sum_{m=1}^\infty\frac{(2m\pi)^2}{(2m\pi)^2+(\kappa L)^2}
  \ln\left(1+\left(\frac{\kappa L}{m\pi}\right)^2\right)
\nonumber
\\ && \qquad
\times \left\{
2\frac{\sinh\frac{\kappa L}{2}\frac{z}{L}
  \sinh\frac{\kappa L}{2}(z/L-1)}{\cosh\frac{\kappa L}{2}}
- 2\left(\frac{\kappa L}{2m\pi}\right)^2
  \sin^2\frac{m\pi z}{L}
\right\}
\nonumber
\\
&=& \frac{8\pi L}{\epsilon^2} 
\Bigg\{
\frac{\sinh\frac{\kappa L}{2}\frac{z}{L}
  \sinh\frac{\kappa L}{2}(z/L-1)}{(\kappa L)^2\cosh\frac{\kappa L}{2}}
\sum_{m=1}^\infty
  \frac{(2m\pi)^2\ln\left(1+\left(\frac{\kappa L}{m\pi}\right)^2\right)}
       {(2m\pi)^2+(\kappa L)^2}
\nonumber
\\ && \qquad
  - \sum_{m=1}^\infty
  \frac{\ln\left(1+\left(\frac{\kappa L}{m\pi}\right)^2\right)}
       {(2m\pi)^2+(\kappa L)^2}
\sin^2\frac{m\pi z}{L}
\Bigg\}
\nonumber
\\
&=& \frac{8\pi L}{\epsilon^2}
\left[\frac{\sinh\frac{\kappa z}{2}
  \sinh\frac{\kappa L}{2}(z/L-1)}{\cosh\frac{\kappa L}{2}}
  I_3(\kappa L)
- I_2(\kappa L,z)
\right]
\end{eqnarray}
\end{widetext}
where $I_2$ and $I_3$ are defined in  Eq.~(\ref{I23definition}).

Finally, for the electric potential, we require the derivative
\begin{eqnarray}
&& \left.\frac{\partial}{\partial z} 
\int dr'\Delta G(r',r') G(r',r) \right|_{z=0}\\
&& \qquad\qquad = -\frac{4\pi\kappa L}{\epsilon^2} 
\tanh\frac{\kappa L}{2} I_3(\kappa L)\nonumber
\end{eqnarray}
which can easily be obtained by differentiating Eq.~(\ref{GGcalc}).

In order to perform asyptotic analysis for the functions $I_1$, $I_2$,
and $I_3$, it is convenient to express them in terms of integrals
rather than an infinite series.  This can be acheived by first noting
that
\begin{eqnarray}
\ln\left(1+\left(\frac{\kappa L}{n\pi}\right)^2\right)
&=& \int_0^{\kappa L} \frac{2xdx}{(n\pi)^2+x^2}
\end{eqnarray}
and then using the relation:
\begin{eqnarray}
\sum_{n=1}^\infty \frac{\cos2m\pi x}{(2m\pi)^2+\alpha^2}
&=& \frac{\cosh\frac{\alpha}{2}(1-2x)\alpha}{4\alpha\sinh(\alpha/2)}
-\frac{1}{2\alpha^2}
\end{eqnarray}
It can then be shown that
\begin{eqnarray}
I_1(\kappa L,z) &=& \frac{(\kappa L)}{2}\int_0^2 dx
\frac{\sinh\frac{\kappa Lx}{2}(1-z/L)\sinh\frac{\kappa xz}{2}}
           {\sinh\frac{\kappa Lx}{2}}
\end{eqnarray}
\begin{eqnarray}
I_2(\kappa L,z) &=& \frac{1}{2\kappa L}\int_0^2 \frac{xdx}{x^2-1}
\Bigg[\frac{\sinh\frac{\kappa L}{2}(1-z/L)\sinh\frac{\kappa z}{2}}
           {\sinh\frac{\kappa L}{2}}
\nonumber
\\ && \qquad
- \frac{\sinh\frac{\kappa Lx}{2}(1-z/L)\sinh\frac{\kappa x z}{2}}
           {x\sinh\frac{\kappa Lx}{2}}\Bigg]
\\
I_3(\kappa L) &=&
\frac{1}{2\kappa L} \int_0^{2}\frac{xdx}{x^2-1}
\nonumber
\\ && \qquad
\times\left(x\coth\frac{\kappa Lx}{2}
  -\coth\frac{\kappa L}{2}\right)
\end{eqnarray}

\section{One-loop Correction}
\label{ZGappendix}
The one-loop correction to the grand potential is given by the expression
\begin{widetext}
\begin{eqnarray}
\delta \ln Z_G[\gamma] 
&=& -\frac{1}{2} {\rm Tr} [\ln(1+\bar{K}_2G_0)-\bar{K}_2G_0]
\nonumber
\\
&=& -\frac{A}{2} \sum_{n=1}^\infty\int\frac{dq_1}{(2\pi)}\frac{dq_2}{(2\pi)}
\left\{
\ln\left(1+\frac{\kappa^2}{(n\pi/L)^2+q_1^2+q_2^2}\right)
- \frac{\kappa^2}{(n\pi/L)^2+q_1^2+q_2^2}
\right\}
= -\frac{\kappa ^2 A}{8\pi} I(\kappa L).
\end{eqnarray}
\end{widetext}
We need to evaluate the following integral
\begin{eqnarray}
\label{I-def:eq}
I(\kappa L) &=&
\sum_{n=1}^\infty\int_0^\infty dx
\left[
\ln\left(1+\frac{1}{(n\pi/\kappa L)^2+x}\right)
\right.
\nonumber
\\ && \qquad
\left.
- \frac{1}{(n\pi/\kappa L)^2+x} \right]
\\ \nonumber
&=& \sum_{n=1}^{\infty}\left\{
1 - \left[ 
\ln \left(1+\left(\frac{\pi n }{\kappa L}\right)^2\right)
-\ln \left(\frac{\pi n }{\kappa L}\right)^2
\right]
\right.
\nonumber
\\ && \qquad
\left.
\times\left[ 1+ \left(\frac{\pi n }{\kappa L}\right)^2 \right]\right\}.
\end{eqnarray}
Now, use the relation
\begin{eqnarray}
\int_0^1dx \ln(1+x/a^2) = (1+a^2)\ln(1+1/a^2)-1
\end{eqnarray}
to find
\begin{eqnarray}\label{krech}
I(\kappa L) &=& - \sum_{n=1}^{\infty}
\int_0^1 dx \ln\left(
1+\frac{x(\kappa L)^2}{\pi^2 n^2}\right)
\nonumber
\\
&=& -\int_0^1dx \ln \left(\frac{\sinh(\sqrt{x} \kappa L)}
{\sqrt{x}\kappa L}\right).
\nonumber
\\
&=& -\frac{2}{(\kappa L)^2}\int_0^{\kappa L}dx 
x\ln \left(\frac{\sinh x }{x}\right).
\end{eqnarray}
Asymptotic limits can easily be obtained from this expression.
In the limit $\kappa L\to0$, we find
\begin{eqnarray}
I(\kappa L) &=& -\frac{2}{(\kappa L)^2}\left[
\frac{(\kappa L)^4}{24}
- \frac{(\kappa L)^4}{1080}
+ \cdots
\right]
\nonumber
\\
&=& -\frac{(\kappa L)^2}{12}
+ \frac{(\kappa L)^4}{540}
+ \cdots
\end{eqnarray}
and 
\begin{eqnarray}
I'(\kappa L) 
&=& - \frac{\kappa L}{6} + \frac{(\kappa L)^3}{135}
+ \cdots
\end{eqnarray}
On the other hand, in the limit $\kappa L\to\infty$, we find
\begin{eqnarray}
I(\kappa L) 
&=&
-\frac{2}{3}(\kappa L)
+ \left(\ln2-\frac{1}{2} + \ln(\kappa L)\right)
\\ \nonumber
&+& \frac{\zeta(3)}{2} (\kappa L)^{-2}
+\cdots
\end{eqnarray}
and
\begin{eqnarray}
I'(\kappa L) &=&
-\frac{2}{3}
+ \frac{1}{\kappa L}
- \frac{\zeta(3)}{(\kappa L)^3}
+ \cdots 
\end{eqnarray}

\end{appendix}


\bibliographystyle{prsty}

\end{document}